\documentstyle[11pt,newpasp,twoside,epsf]{article}
\def\edcomment#1{\iffalse\marginpar{\raggedright\sl#1\/}\else\relax\fi}
\reversemarginpar

\begin{document}
\title{Migration and Accretion of Protoplanets in 2D and 3D
        Global Hydrodynamical Simulations}
   \author{Gennaro D'Angelo and Willy Kley}
\affil{Computational Physics,
        Auf der Morgenstelle 10,
        D-72076 T\"ubingen, Germany}
\author{Thomas Henning}
\affil{Max-Planck-Institut f\"ur Astronomie,
        K\"onigstuhl 17,
        D-69121 Heidelberg,
        Germany}

\begin{abstract}
Planet evolution is tightly connected to the dynamics of both distant
and close disk material. Hence, an appropriate description of disk-planet
interaction requires global and high resolution computations, which
we accomplish by applying a Nested-Grid method. Through simulations in
two and three dimensions, we investigate how migration and accretion
are affected by long and short range interactions. For small mass objects,
3D models provide longer growth and migration time scales than 2D ones do,
whereas time lengths are comparable for large mass planets.
\end{abstract}
\section{Introduction}
Migration has entered the puzzling scenario of planetary formation as
the favorite mechanism advocated to explain the extremely short orbital
period of many extrasolar planets.

It is known that any planet-like body is forced to adjust its distance
from the central star because of gravitational interactions with the
circumstellar material.  However, the dispute about how fast migration
proceeds is far from being over.  Numerical methods have been employed
to evaluate gravitational torques exerted on embedded planets. We have
performed a series of simulations modeling both two and three
dimensional disks, varying the mass $M_p$ of the protoplanet in the
range from 1 Earth-mass to 1 Jupiter-mass.

The physics of the problem demands that the flow in the protoplanet's
neighborhood should be accurately resolved. In order to achieve sufficient
resolution, even for very small planetary masses, we use a
\textit{Nested-Grid} technique (D'Angelo, Henning, \& Kley 2002a).
This paper addresses the issues of flow circulation around protoplanets,
orbital migration, and mass accretion.
\section{Numerical Model}
We assume the protostellar disk to be a viscous fluid
(viscosity $\nu=10^{15}\;\mathrm{cm}^2\,\mathrm{s}^{-1}$) and describe it
through the Navier-Stokes equations (Kley, D'Angelo, \& Henning 2001;
D'Angelo et al.\ 2002a). The set of equations is integrated over a
grid hierarchy, as shown in Figure~1. The planet is supposed to move
on a circular orbit at $r_p=5.2\;\mathrm{AU}$ around a solar-mass star.
The disk has an aspect ratio $h=0.05$ and the mass within the simulated
region is $M_D=3.5\times 10^{-3}\;M_{\sun}$.
\begin{figure}
\begin{minipage}[c]{0.4\textwidth}
\plotone{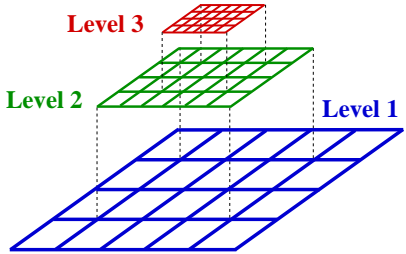}
\end{minipage}\hfill%
\begin{minipage}[c]{0.6\textwidth}
\caption{%
Sketch of a 2D, three-level Nested-Grid system. Each of
the first two grid levels hosts a finer grid, allowing for an
increasing accuracy. In these computations up to 7 grid levels
were employed.}
\end{minipage}\hspace*{\fill}
\end{figure}
\section{2D Simulations of Uranus and the Earth}
A circumplanetary disk forms inside the Roche lobe of massive as well as
low-mass protoplanets. Such structures are characterized by a two-arm spiral
wave perturbation. They are detached from the circumstellar disk spirals,
which arise outside of the planet's Roche lobe.
Indicating with $l$ the distance from the planet
normalized to $r_p$, for a wide range of planetary masses the spiral pattern
can be approximated to
\begin{equation}
\Theta-\Theta_0 = 2\,k\left(%
                  1/\sqrt{l_0} -
                  1/\sqrt{l}\right),
\end{equation}
where $k=\zeta\,\sqrt{M_p/M_*}/h$ and $\zeta\approx 1$.
The ratio $k/\sqrt{l}$ represents the Mach number of the
circumplanetary flow. Figure~2 (left panel) demonstrates how equation~(1)
fits to the spiral perturbation around a Uranus-mass planet.

Even an Earth-mass planet induces two weak spirals, which wrap
a\-round the star for $2\pi$, but no circumplanetary disk is observed
(Figure~2, right panel).

\begin{figure}
\centerline{\epsfxsize=0.428\textwidth \epsfbox{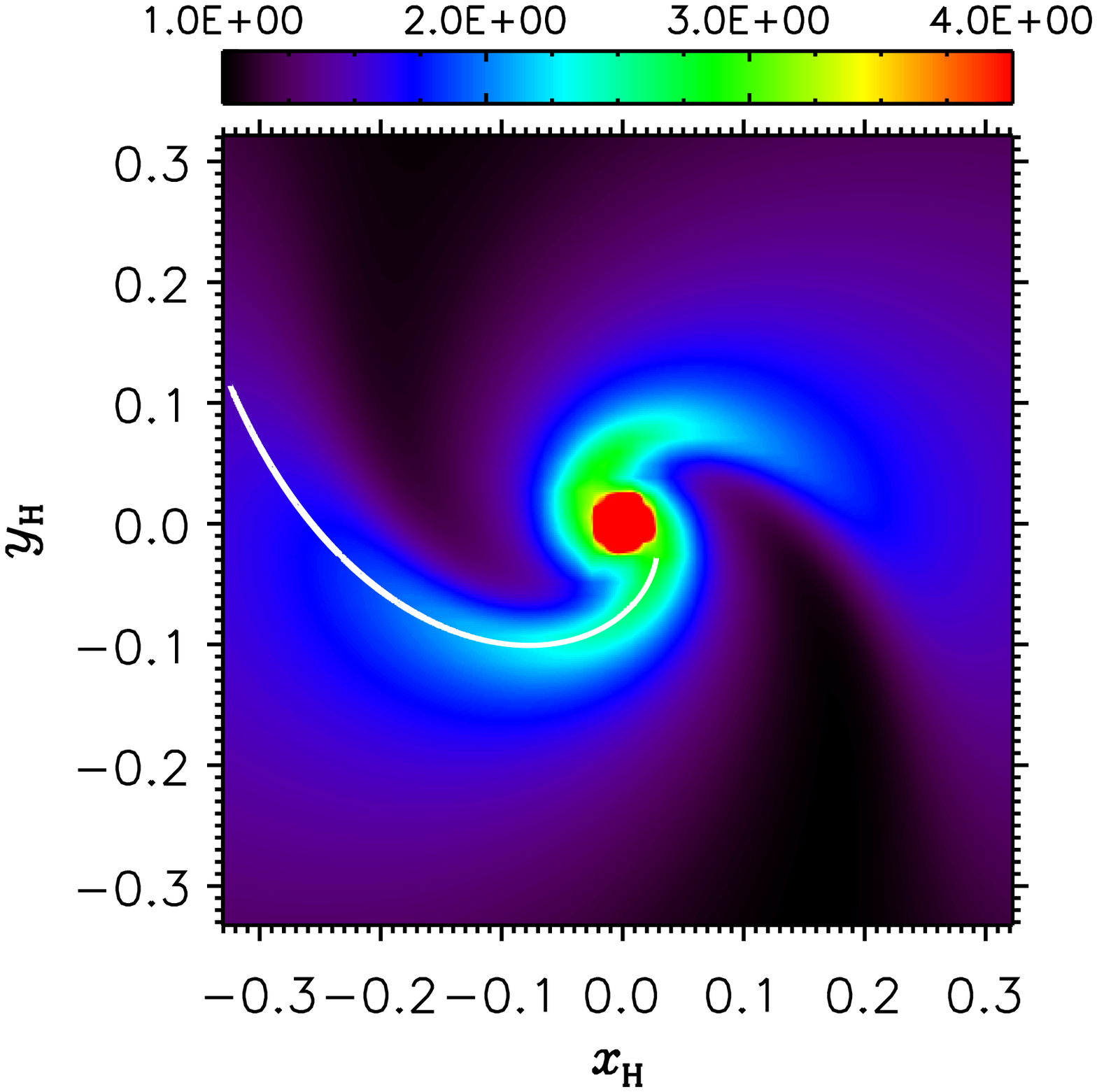}%
         \epsfxsize=0.428\textwidth \epsfbox{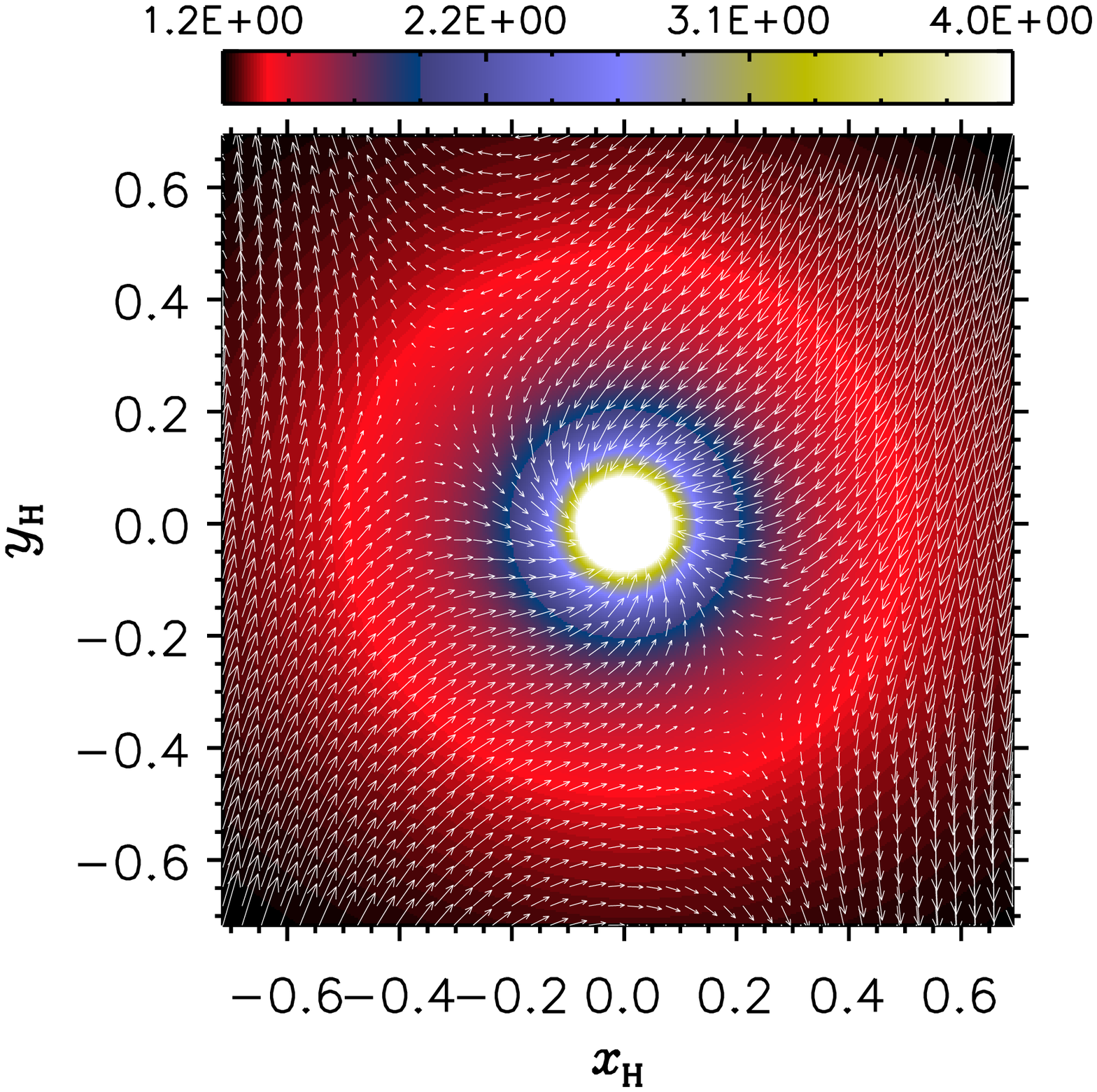}}
\caption{%
Surface density around a Uranus (\textit{left}) and
an Earth (\textit{right}) mass protoplanet. Axis scales are in Hill
coordinates. In physical units, $\Sigma=4$ corresponds to
$256\;\mathrm{g}\,\mathrm{cm}^{-2}$.}
\end{figure}
\section{3D Simulations of Saturn and Neptune}
\begin{figure}
\centerline{\epsfxsize=0.45\textwidth \epsfbox{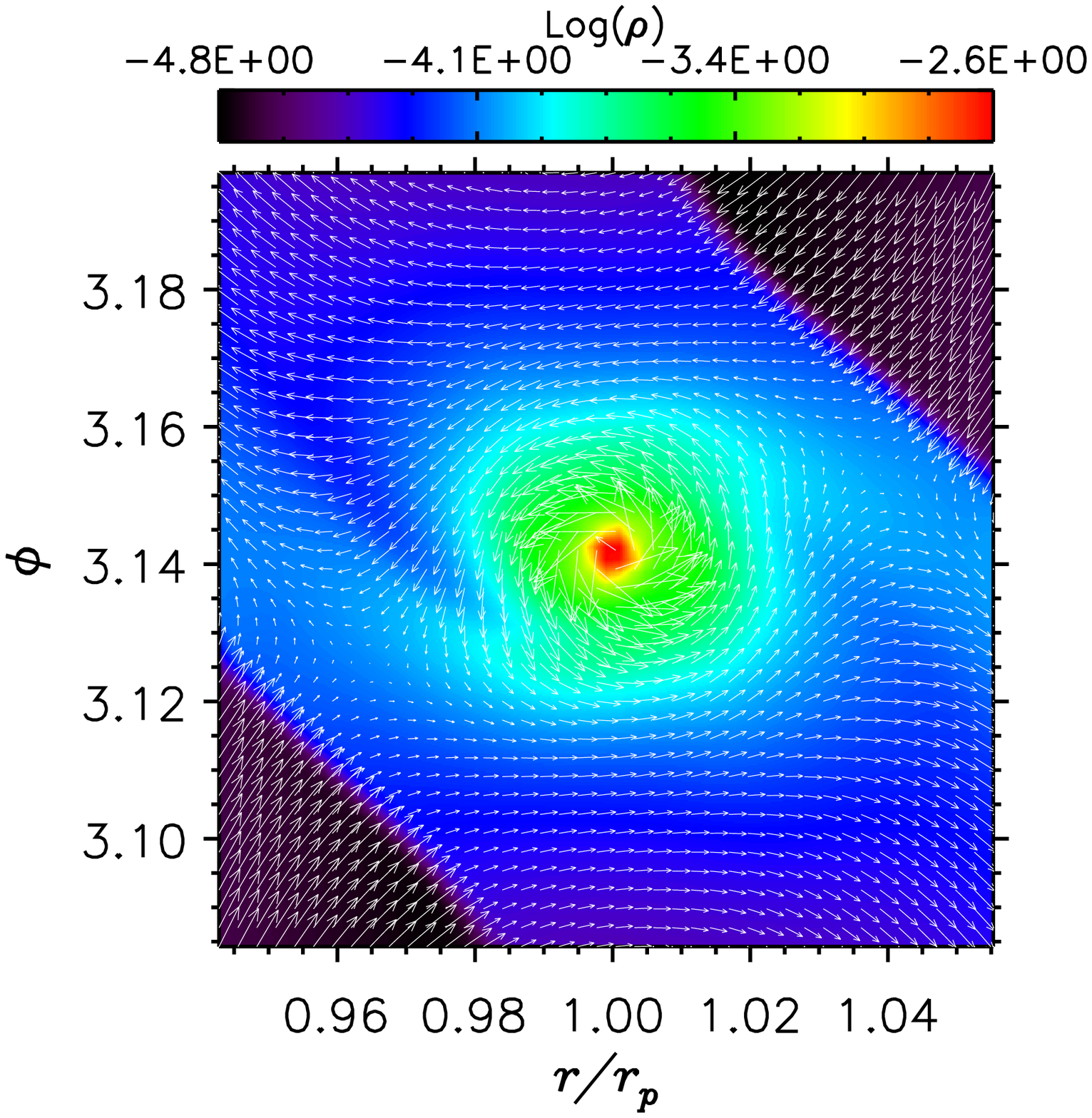}%
         \epsfxsize=0.45\textwidth \epsfbox{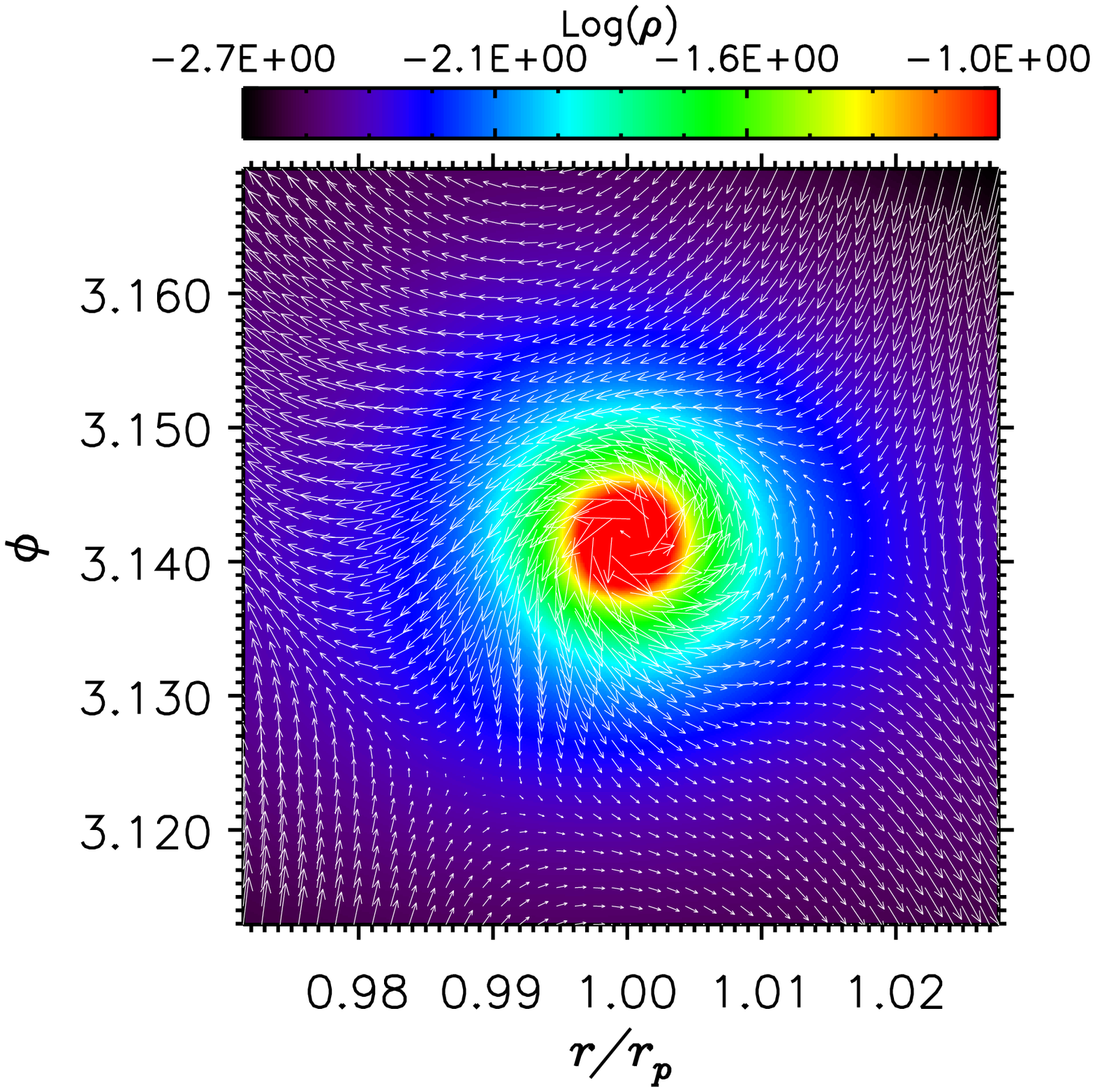}}\vspace*{2mm}
\centerline{\epsfxsize=0.45\textwidth \epsfbox{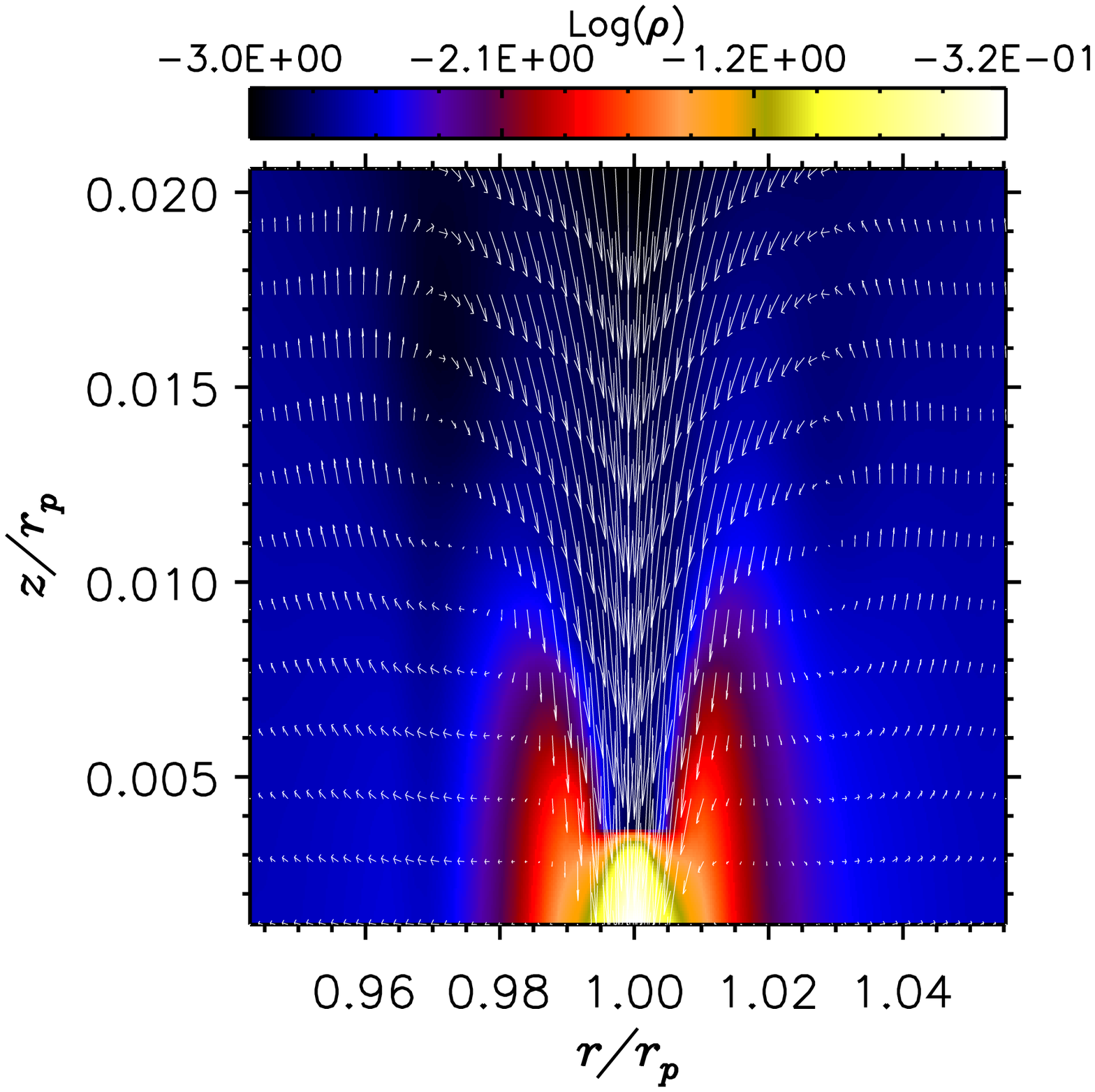}%
         \epsfxsize=0.45\textwidth \epsfbox{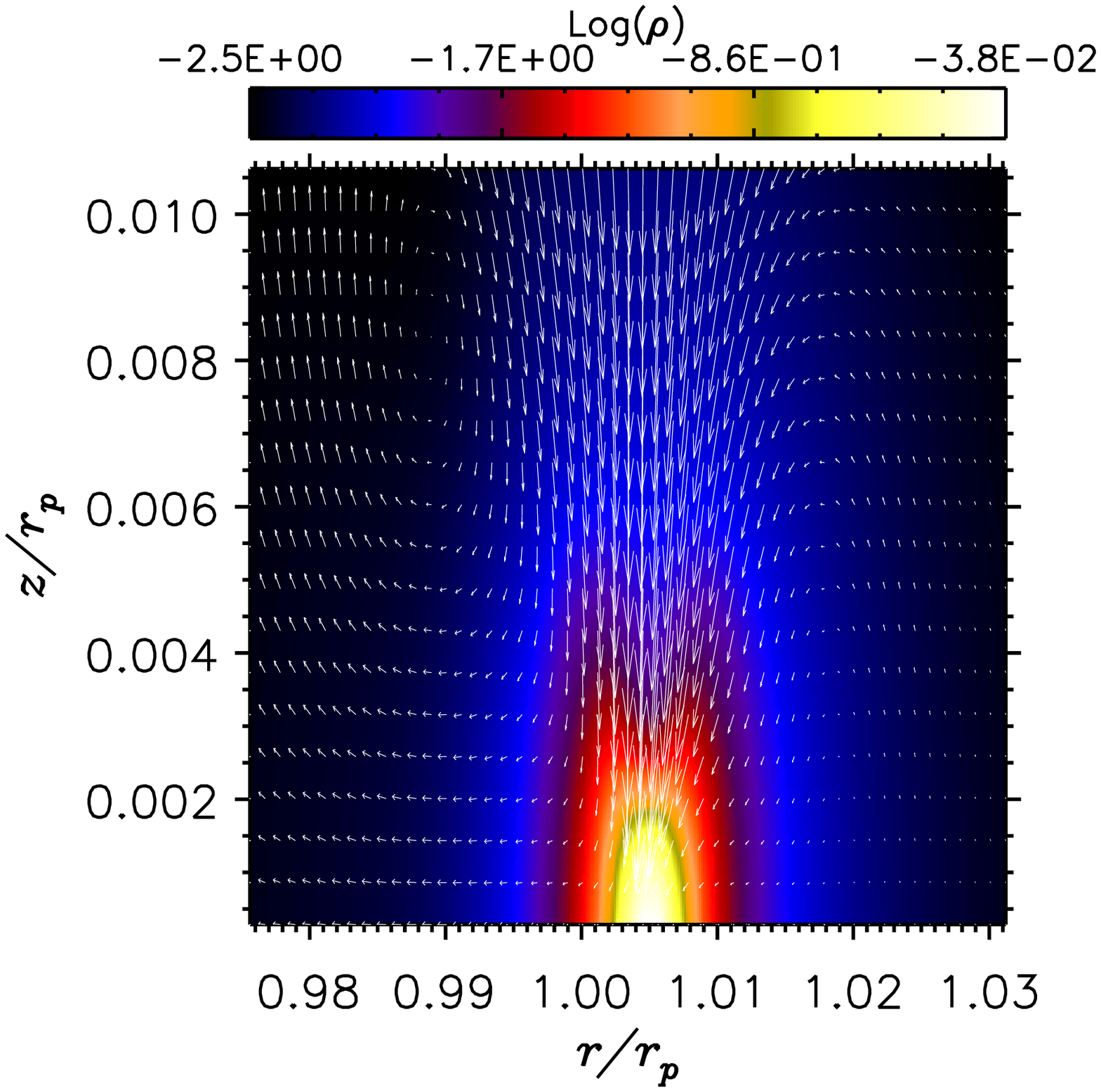}}
\caption{%
Density and velocity field in the equatorial plane (\textit{top}) and
in a vertical plane containing the planet (\textit{bottom}).
\textit{Left panels} refer to Saturn, \textit{right panels} to Neptune.
In the plot, $\rho=0.1$ corresponds to
$4.2\times 10^{-10}\;\mathrm{g}\,\mathrm{cm}^{-3}$.
Maximum velocities are on the order of $3\;\mathrm{km}\,\mathrm{s}^{-1}$.}
\end{figure}
\begin{figure}
\plottwo{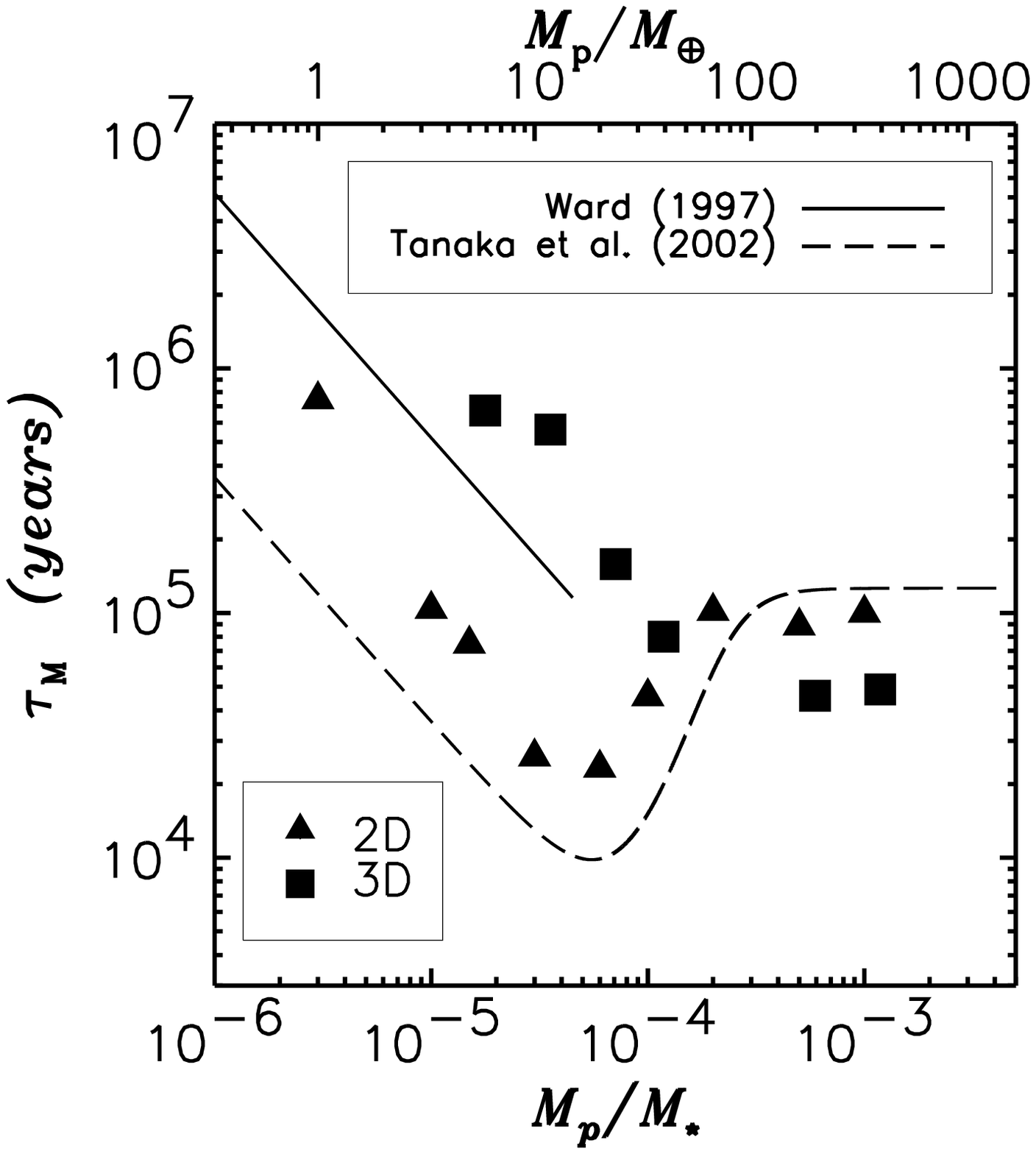}{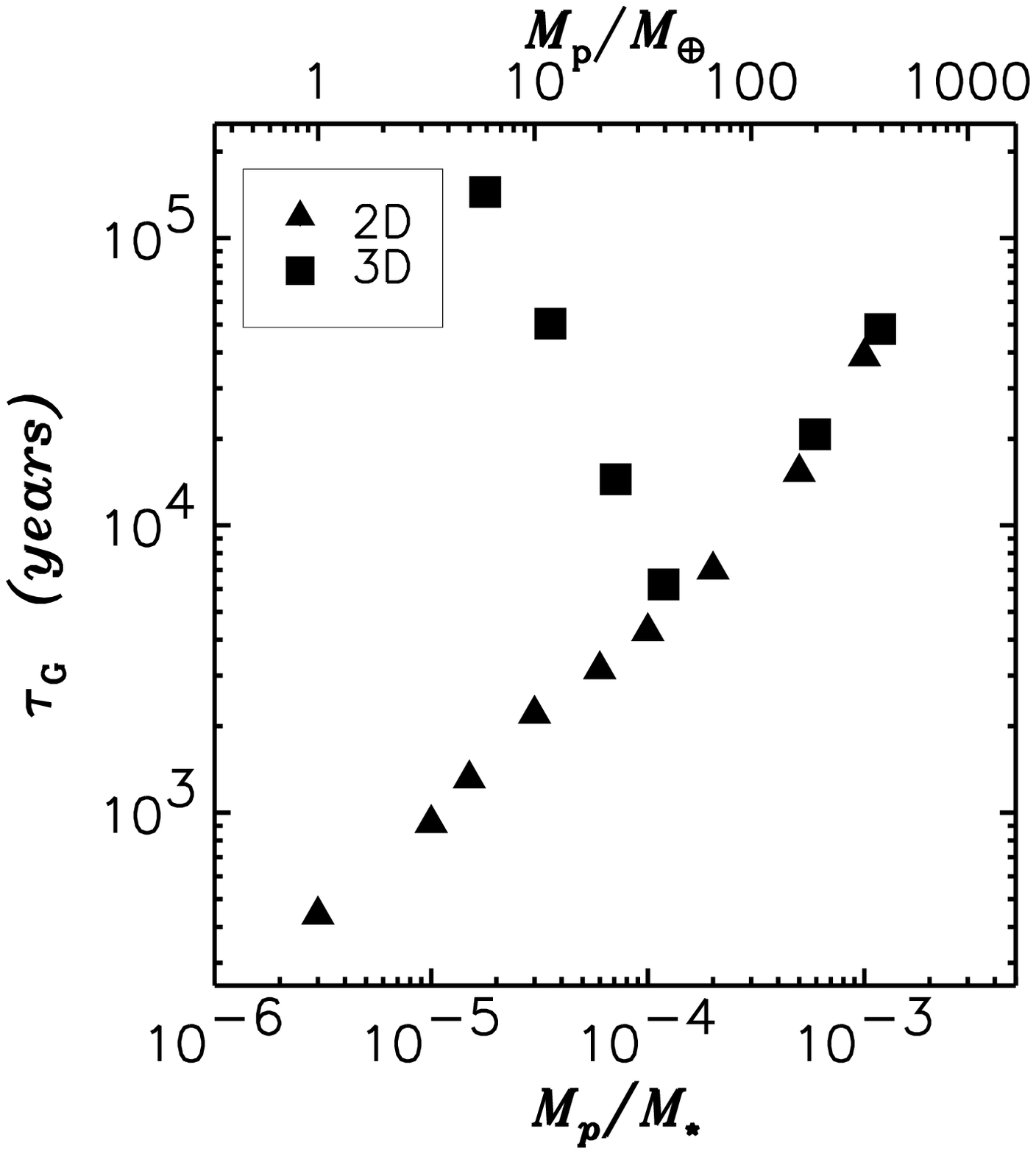}
\caption{%
Migration (\textit{left}) and growth (\textit{right}) time scale
for different planetary masses. Migration is compared to two analytical
models: Ward's one doesn't include corotation torques whereas the
other does.}
\end{figure}

A more complete description of the flow near protoplanets is provided by 3D
computations (D'Angelo, Kley, \& Henning 2002b). The major differences between
2D and 3D modeling arise in the vicinity of the planet because the latter
can account for the vertical circulation in the circumplanetary disk.
Instead, the two geometries decently agree on length scales larger than the
disk scale height.

Two examples of our simulations are illustrated in Figure~3. The
images represent the logarithm of the density close to a Saturn-mass
planet (left) and Neptune-mass planet (right), in two orthogonal
planes (see Figure caption for details).  The velocity field is
overplotted to display the flow features. From the top panels it is
clear that the spiral perturbations are weaker and more open than in
2D simulations. The bottom panels indicate the presence of vertical
shock fronts, which are generally located outside the Hill sphere of
the protoplanet.

\section{Migration and Accretion}

In general, torques exerted by disk material cause a protoplanet to migrate
toward the star (Ward 1997). Yet, nearby matter can be very efficient at
slowing down its inward motion (Tanaka, Takeuchi, \& Ward 2002; D'Angelo et
al.\ 2002a). The migration time scale can be defined as
$\tau_{\mathrm{M}}\equiv r_p/|\dot{r}_p|$, where the migration drift
$\dot{r}_p$ is directly proportional to the total torque acting on the planet.
Since we also measure the rate $\dot{M}_p$ at which the planet accretes
matter from its surroundings, an accretion time scale can be introduced as
well: $\tau_{\mathrm{G}}\equiv M_p/\dot{M}_p$.

Some of our 2D and 3D outcomes for both time scales are shown in Figure~4.

\section{Summary}
Circumplanetary disk forms around protoplanets. The spiral wave pattern
which marks such disks is less accentuated when the full 3D structure is
simulated. Vertical shock fronts develop outside the Hill sphere of
the planet.

The estimated values of $\tau_{\mathrm{M}}$ in 3D are longer than those
predicted by analytical linear theories because of non-linearity effects.
When $M_p\la 30\;M_{\earth}$, both $\tau_{\mathrm{M}}$ and
$\tau_{\mathrm{G}}$ are longer in 3D computations then they are in 2D
ones.

\end{document}